\documentclass{article}
\usepackage{graphicx} 
\usepackage{amssymb}
\usepackage{amsmath}
\usepackage{amsthm}
\usepackage{color}
\usepackage{comment}
\usepackage{enumitem}		
\usepackage{geometry}		
\usepackage{graphicx}
\usepackage{authblk}
\usepackage{amssymb}
\usepackage{listings}
\usepackage{xcolor}
\usepackage{caption}
\usepackage{color}
\usepackage{algorithm}
\usepackage{algpseudocode}
\usepackage{subcaption}
\usepackage{hyperref}
\usepackage{url}
\title{Spatiotemporal patterns in a 2D lattice of Hindmarsh-Rose neurons induced by high-amplitude pulses}
\author[a]{J.S. Ram}
\author[a]{S.S. Muni}
\author[b,c]{I.A. Shepelev}
\affil[a]{School of Digital Sciences, Digital University Kerala, Technopark phase-IV Campus, Mangalapuram, 695317, Thiruvananthapuram, India}
\affil[b]{Almetyevsk State Petroleum Institute, Lenin St., 2, Almetyevsk 423462, Russia}
\affil[c]{Saratov State University, 83 Astrakhanskaya Street, Saratov 410012, Russia}
\graphicspath{{figures/}}
\usepackage[english]{babel}

\begin{document}

\maketitle
\begin{abstract}
We present numerical results for the effects of influence by high-amplitude periodic pulse series on a network of nonlocally coupled Hindmarsh-Rose neurons with 2D geometry of the topology. We consider the case when the pulse amplitude is larger than the amplitude of oscillations in the autonomous network for a wide range of pulse frequencies. An initial regime in the network is a spiral wave chimera. We show that the effects of external influence strongly depend on a balance between the pulse frequency and frequencies of the spectral peaks of the autonomous network. Except for the destructive role of the pulses, when they lead to loss of stability of the initial regime, we have also revealed a constructive role. We have found for the first time the emergence of a new type of multi-front spiral waves, when the wavefront represents a set of several close fronts, and the wave dynamics are significantly different from common spiral waves: neurons oscillate independently to the wave rotation, the rotation velocity is in many times less than for the common spiral wave, etc. We have also discovered several types of cluster spatiotemporal structures induced by the pulses.
\end{abstract}

\section*{Introduction}
Many natural phenomena and technological constructs depict networks composed of interacting nonlinear elements \cite{Strogatz2001, Yang2023, Gosak2018, Havlin2012}. They can be populations of living organisms, power grids \cite{Pagani2013}, traffic flow in transportation \cite{Rocha2017}, social networks \cite{Bedi2016}, neural networks in the brain \cite{Chialvo2010, Telesford2011} and many more. Examining such systems is currently a rapidly expanding field within nonlinear dynamics. Neurons are the structural and functional building blocks of the nervous system, where the transmission of information happens through chemical and electrical signals. As a nonlinear dynamical system, neurons display excitability because they are near a transition, called bifurcation, from resting to continuous spiking activity. The transition is due to potential differences between the inner and outer surfaces of the neural membrane. The neural dynamic systems were typically characterized by mathematical models that included nonlinear ordinary differential equations or discrete maps, for example, Hodgkin-Huxley neuron \cite{hodgkin1952quantitative}, FitzHugh-Nagumo neuron \cite{FitzHugh1961,nagumo1962active}, Izhikevich neuron\cite{izhikevich2003simple}, and Hindmarsh-Rose neuron \cite{1984}. Excitable media serve as a framework where complex spatiotemporal patterns can arise \cite{Bittihn2017, Meron1992}. \textcolor{black}{The presence of a traveling chimera state in a two-dimensional network of HR neurons was studied in works like \cite{Simo2021, GRSIMO2021}. The influence of the weak external electric field on HR neurons by local electrical coupling and nonlocal chemical coupling was studied in the latter. They found traveling multicluster chimera breathers and multicluster chimera breathers in the absence of an electric field and local electrical coupling. In the presence of an electric field on the network, alternating chimera states were observed}. Spiral waves, intriguing spatiotemporal patterns, are commonly found in two-dimensional lattice networks of neurons \cite{parastesh2019birth, Bukh2019}. The chemical Belousov-Zhabotinsky reaction \cite{ZAIKIN1970}, propagating electrical activity in heart tissue \cite{Gray1998}, calcium waves within Xenopus oocyte cells \cite{Lechleiter1991}, calcium waves between cells in brain tissue samples (specifically in hippocampal slice cultures) \cite{HarrisWhite1998}, and the oxidation of CO on platinum (Pt110) surfaces \cite{Br1994} are examples of real systems that exhibit spiral waves. The emergence of spiral wave regimes in biological systems has harmful effects. For instance, spiral waves in cardiac tissues disrupt normal rhythmic activities, leading to arrhythmia \cite{GRAY1996}. Therefore, studying spiral waves not only aids in understanding their mechanism but also holds the key to their control.

In computational neuroscience, mathematical models are essential for understanding the nuances of neural dynamics. The Hindmarsh-Rose neuron model is an elementary framework replicating neuron's functioning. The model exhibits various neuronal behaviors, including regular spiking, bursting, and chaotic activity, depending on the parameter values \cite{SHILNIKOV2003}. Researchers have extensively investigated the Hindmarsh-Rose neuron's role in diverse network topologies, and each offers unique insights into the collective dynamics of interconnected neurons. A Bandyopadhyay et al. \cite{Bandyopadhyay2018} explores how structural properties of different networks affect the synchronization of HR neurons coupled by sigmoid coupling function. They have illustrated a comparison of synchronizability for complex topologies like small-world \cite{PhysRevLett.89.054101, Bassett2006}, scale-free \cite{Batista2007}, and modular \cite{Park2006} networks. The existence of double-well chimera states in both the ring-star \cite{Muni2020} and ring \cite{Kriener2014} networks using the memristive HR neuron model has been mentioned for the first time in recent research by one of our co-authors \cite{MUN2022}. The formation of non-stationary chimera patterns in three-dimensional locally coupled HR neurons is studied in \cite{Kundu2019}. Another research discusses the robustness and the mechanisms leading to the breakdown of spiral waves in a two-dimensional lattice network of HR neurons \cite{Ma2010}. In Ref. \cite{Parastesh2022}, authors examine the role of a global coupling scheme in achieving synchronization within a simplicial complex of Hindmarsh-Rose neurons that engage in higher-order interactions \cite{Benson2016, Battiston2021}. Electrical gap junctions define first-order interactions. Nonlinear chemical synaptic couplings and linear diffusive couplings, which resemble electrical synapses, are the two circumstances under which second-order (or three-body) interactions are considered. The extensive research on the Hindmarsh-Rose neuron model in different network topologies provides valuable insights into synchronization, spatiotemporal patterns, and the behavior of complex neural networks.

Noise is inevitable in real neuronal systems. Intuitively, we consider random fluctuations or noise to have a disruptive effect. However, several findings from theoretical analysis \cite{Wiesenfeld1994,Hnggi2002}, numerical simulations \cite{McDonnell2011}, and experimental studies \cite{Braun1994,Stacey2001} have shown the constructive impact of noise, for example, effects of stochastic resonance \cite{ando2000stochastic} or coherence resonance \cite{Pikovsky1997}. As for spiral waves,  it has been shown that noise influence can play a crucial role in the formation or destruction of spiral waves in neuronal networks \cite{Kang2020, Song2023}.

Real-world oscillatory systems often encounter impulsive forces rather than continuous inputs from diverse sources such as neural activity, mechanical shocks, or electrical pulses \cite{Glass2001}. Recent studies have focused on the influence of external periodic impulses on a network of oscillators. Controlling chaos in complex networks has been a significant research topic due to its implications for various fields, including biological systems, neural networks, and engineering applications. The application of carefully selected periodic external stimulations has been proven effective in controlling chaos in various interconnected systems like arrays of electrochemical oscillators \cite{Wang2001}, Frenkel-Kontorova chains \cite{Martnez2004}, and recurrent neural networks \cite{Rajan2010}. In \cite{HAN2008}, the authors propose a novel coupled model where synchronization is achieved by using impulsive control strategies at specific moments rather than through continuous coupling. This approach reflects real-world scenarios where instantaneous phenomena, such as population growth, spacecraft manoeuvres, and dynamical nerve cell networks, are best modeled by impulsive differential equations. Further, numerical simulations of the theoretical results were carried out on typical chaotic systems like Chua's circuit and the Lorentz system. In the research paper \cite{Chacn2016}, R. Chacón and colleagues conducted a study on controlling chaos within star-like networks of damped kicked rotators composed of dissipative nonlinear oscillators. They emphasize a method for reducing chaotic behavior by diminishing the impact of periodic pulses at a local level. In a recent study \cite{MOREIRA2019487}, Moreira and Aguiar explored the global synchronization of Kuramoto oscillators on networks with partial external forcing. Their findings indicated that synchronization is easier to attain with a higher number of nodes being driven externally, but there is a critical threshold below which synchronization cannot be achieved.

The works mentioned above have motivated us to study the impact of Gaussian pulses on a 2D lattice of Hindmarsh-Rose neurons. Our research extends upon this groundwork by showing how regular Gaussian pulses can disrupt current patterns and generate novel, reliable dynamic behaviors in these networks, emphasizing the practical importance and possible uses of impulsive control methods in real-life oscillatory systems. The rest of the paper is organized as follows: In \S 1, we introduce the lattice network of the two-dimensional Hindmarsh-Rose neuron model. \S 2 delves into the dynamics of the system under the external periodic pulse force. Novel spatiotemporal patterns were seen as a result of the application of a strong Gaussian pulse. Further, we explain the characteristics of each region in the regime map in detail. Finally, the paper concludes with a conclusion and future direction.

\section{Network equations}

The Hindmarsh-Rose neuron model is a basic system that simulates neuronal activity. The model demonstrates a range of neuronal activities, such as consistent spiking, bursting, and chaotic behavior, which vary based on the values of its parameters. It is characterized by a system of 3 ODE's as shown in \cite{1984}.
\begin{equation}
\begin{aligned}
\dot{x}=~&y-ax^3+bx^2-z+I_{ext}, \\
\dot{y}=~&c-dx^2-y, \\
\dot{z}=~&r[s(x-\chi)-z]
\end{aligned}
\label{eq:vdP}
\end{equation}
where variable $x$ denotes the difference of electrical potentials across the neuron’s membrane, whereas the variable $y$ signifies the fast current (Na$^+  $ or K$^+ $ ions) acting as a recovery variable. The variable $z$ represents a slow adaptation current, primarily involving Ca$^{2+}$ ions.
\textcolor{black}{The parameter $a$ is a recovery constant, $b$ enables us to switch between bursting and spiking regimes and to control the spike frequency, $I$ corresponds to the input membrane current, the parameter $c$ represents the constant input current. $d$ determines the rate of recovery after spiking, $r$ controls the variation speed of the variable $z$,  $s$ influences the spiking behavior of the neuron without accommodation and subthreshold adaptation, $\chi$ sets the resting potential of the neuron \cite{1984}}.

In turn, a network of Hindmarsh-Rose neurons with 2D geometric topology is described by the following system of ordinary differential equations: 
%
\begin{equation}
\begin{array}{l}
\dot{x}_{i,j} = y_{i,j}-ax_{i,j}^3+bx_{i,j}^2-z_{i,j}+I_{ext} + F(t)
+\\[10pt]
~~~~~~+\dfrac{\sigma}{Q}
\sum\limits_{m,n} \left(x_{m,n} - x_{i,j} \right),
\\[10pt]
\dot{y}_{i,j} = c-dx_{i,j}^2-y_{i,j},\\
\dot{z}_{i,j}= r[s(x_{i,j}-\chi)-z_{i,j}]
\end{array}
\label{eq:grid}
\end{equation}

where ($x_{i,j},\,y_{i,j},\,z_{i,j}$) are dynamical variables, with dual low indices $(i,j)$, indicating the element's location in a two-dimensional lattice, where $i,j = 1, \dots, N=50$. The coefficient $\sigma $ corresponds to the coupling strength. The model equations Eq.~\eqref{eq:grid} are integrated using the Runge-Kutta 4th order method with the time step $dt = 0.005$.

\begin{figure}[!h]
\centering
\includegraphics[width=0.85\linewidth]{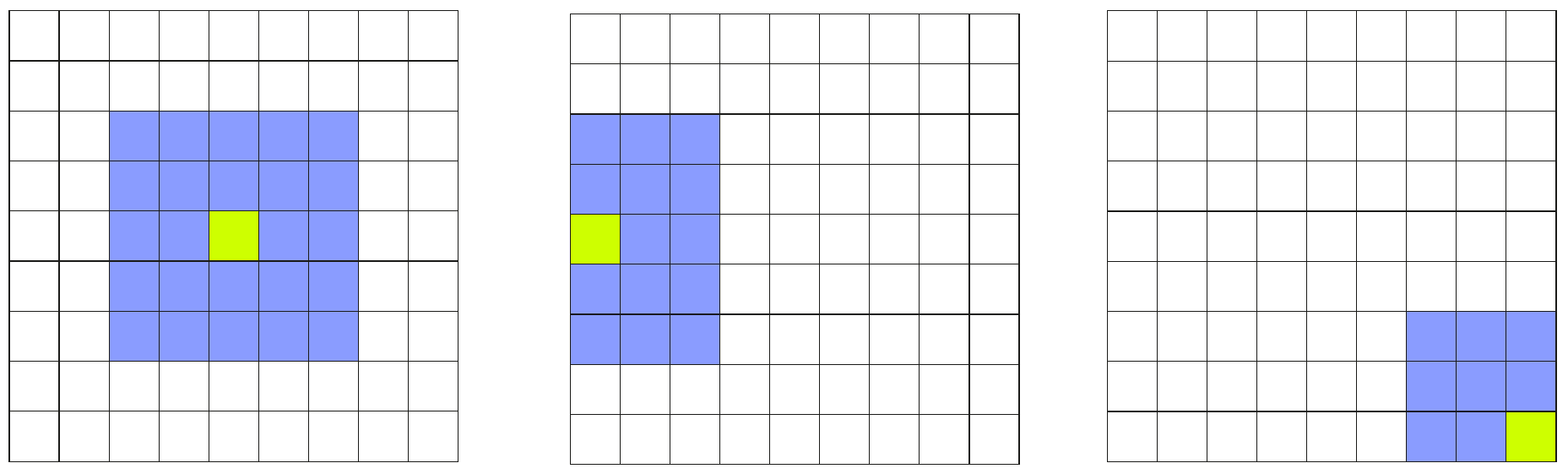}\\
(a)~~~~~~~~~~~~~~~~~~~~~~~~~~~~~~~~~~ (b) ~~~~~~~~~~~~~~~~~~~~~~~~~~~~~~~~~~(c)
\caption{ \textcolor{black}{Schematic representation of the nonlocal coupling with no-flux boundary conditions Eq.~\eqref{eq:free}. Schemes of the coupling in Eq.~\eqref{eq:grid} for different locations of the selected node (marked in yellow): (a) in the lattice center, (b) in the middle of the left edge, and (d) in the right bottom corner. Oscillators coupled with a selected yellow ($i,~ j$)th node are marked in blue, and the remaining uncoupled nodes are shown in white}.}
\label{fig:no-flux}
\end{figure}

The coupling term is included in the initial equation of the lattice as presented in Eq.~\eqref{eq:grid}. This method of incorporating links within the layer amounts to attractive resistive coupling in the second layer of the lattice and repulsive active coupling in the first layer. The coupling strength is denoted by $P$, which indicates the coupling range sizes. The number $Q$ represents the total number of links in both directions for each node, calculated based on the coupling range value $P$. \textcolor{black}{For each node, $Q$ is a measure that combines all links defined by the indices $m$ and $n$, in accordance with the relations that satisfy the no-flux boundary conditions:}   
\begin{equation}\label{eq:free}\left\{\begin{aligned}
&\max(1,i-P) \leqslant m_l \leqslant \min(N,i+P), \\
&\max(1,j-P) \leqslant n_l \leqslant \min(N,j+P),
\end{aligned}\right.\end{equation}
\textcolor{black}{This type of the boundary conditions is shown in Fig.~\ref{fig:no-flux} for different locations of the $i,\,j$th oscillator when the coupling range $P=2$}. 

The external force   is described by the following expression:
\begin{equation}
\begin{aligned}
F(t) = A \,\ exp\left[-\dfrac{sin^2(\frac \omega 2 t)}{2 s}\right]
\end{aligned}
\label{eq:Gauss}
\end{equation}
\begin{figure}[!t]
\centering
\hspace{-2mm}\parbox[c]{.35\linewidth}{ 
  \includegraphics[width=\linewidth]{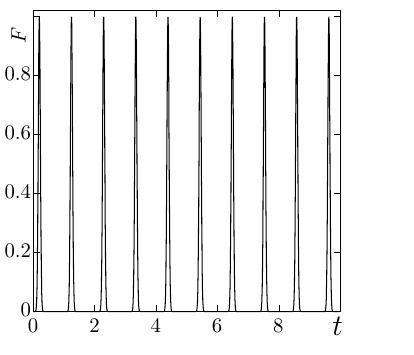}
    \vspace{-7.5mm} \center (a)
}
\hspace{-1mm}\parbox[c]{.35\linewidth}{
\includegraphics[width=\linewidth]{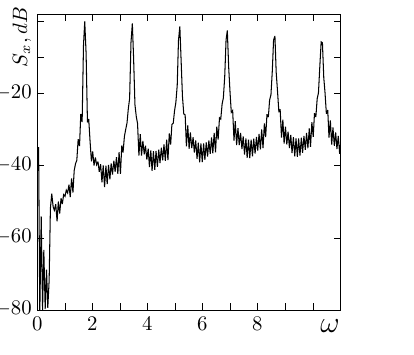}
\vspace{-7.5mm}\center (b)
}
\caption{ (a) Periodic series of Gaussian pulses described by Eq.~\eqref{eq:Gauss} for $A = 1$, $\omega = 3$, $s =0.01$ and their corresponding normalizing spectral density (b).
 }
\label{fig:GaussianForce}
\end{figure}
It represents a periodic time series of Gaussian pulses \cite{chomycz2009gaussian,Bauer1984} with an amplitude $A$, which periodically repeats in time with the frequency $\omega$. We use Gaussian pulses as the external force in order to maintain the integrability of the system. The width of the pulse is determined by the parameter $s$. Within the study, a value of $s$ is fixed and equal to $s = 0.01$. An example of a time series for the external force is shown in Fig.~\ref{fig:GaussianForce}(a). Fig.~\ref{fig:GaussianForce}(b) shows the corresponding normalised spectral density.
Note that the real frequency of the pulses is equal to doubled $\omega$. For this reason, we divide the expression by 2 in \eqref{eq:Gauss}. The function \eqref{eq:Gauss} is a continuously differentiable function, unlike the delta function. At the same time, the pulse duration is very short. Hence, the influence is very short-term. 

\begin{figure}[!b]
\centering
\hspace{-2mm}\parbox[c]{.33\linewidth}{ 
  \includegraphics[width=\linewidth]{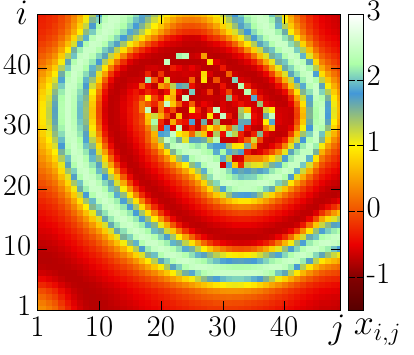}
    \vspace{-7.5mm} \center {(a), $t=100$}
}
\hspace{-1mm}\parbox[c]{.33\linewidth}{
\includegraphics[width=\linewidth]{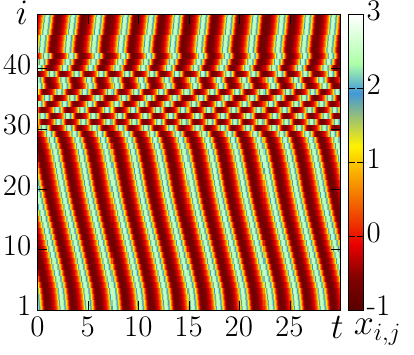}
\vspace{-7.5mm}\center (b), $j=26$
}
\hspace{-1mm}\parbox[c]{.33\linewidth}{
\includegraphics[width=\linewidth]{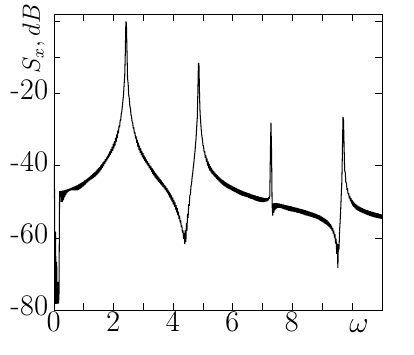}
\vspace{-7.5mm}\center (c)
}
\caption{(Color online) Initial regime in the autonomous network Eq.~\eqref{eq:grid} ($A = 0$) for $\sigma=0.145 $ and $P = 2$. (a) is a snapshot of the system state, (b) presents a space-time plot of the cross-section $j=25 $. (c) is a normilized spectral density for $i = 30,\, j=25$. Parameters: $a=1,\,b=3,\,c=1,\,d=5,\,r=0.006,\,s=4,\,\chi=1.6 $, $N=50$.} 
\label{fig:initial_regime}
\end{figure}

\textcolor{black}{An initial regime in the autonomous system \eqref{eq:grid} ($A=0$) represents a spiral wave chimera, which is depicted in Fig.~\ref{fig:initial_regime}(a). It remains without both qualitative and quantitative changes in its features during the whole observation time (20000-time units)}. \textcolor{black}{It occurs for randomly distributed initial conditions within $x_{i,\,j} \in [-0.95, 1.05]$, $y_{i,\,j} \in [-0.98, 1.02]$, $z_{i,\,j} \in [-0.97, 1.03]$ and for $P=2$, $\sigma = 0.145$} after a transient time $T_{\rm tr} = 10000$ time units.  This wave structure is typical for nonlocally coupled network \eqref{eq:grid}. The rotating spiral wavefront encircles the core of incoherence where the instantaneous phases of adjacent oscillators are significantly different. A space-time diagram of the cross-section shows the spatiotemporal dynamics through the incoherence core in Fig.~\ref{fig:initial_regime}(b) at $j=25$. The core localization is stable in time and does not change the spatial range. The dynamics of the rest part of the network are determined by the wave rotation. Hence, the mean frequency of all the nodes is the same and equal to $\approx 2.25$. The oscillations have a nonlinear relaxation character.

Their normalized power spectral density is illustrated in Fig.~\ref{fig:initial_regime}(c). The spectrum consists of four spectral maxima within $\omega\in [0,11]$. The main harmonic has frequency $\omega_{max} = 2.26$ and is almost equal to the mean frequency. \textcolor{black}{We use the instantaneous system state after 10000-time units of observation as initial conditions for all the following studies}. 

\section{Results}
\subsection{Regime map}
\begin{figure}[!h]
\hspace{-1mm}\parbox[c]{.7\linewidth}{ 
  \includegraphics[width=\linewidth]{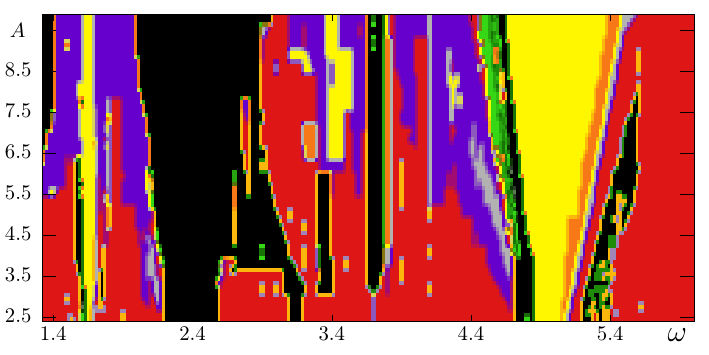}
    \vspace{-7.5mm} \center (a)
  \includegraphics[width=\linewidth]{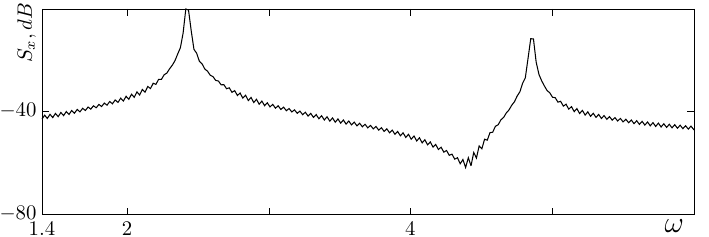}
    \vspace{-7.5mm} \center (c)
}
\hspace{-1mm}\parbox[c]{.234\linewidth}{
\includegraphics[width=\linewidth]{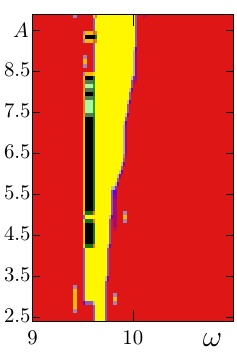}
\vspace{-7.5mm}\center (b)
\includegraphics[width=\linewidth]{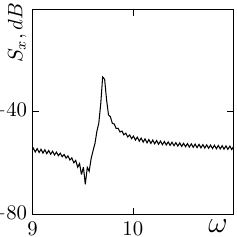}
 \vspace{-7.5mm} \center (d)
}
\centering
 \includegraphics[width=.7\linewidth]{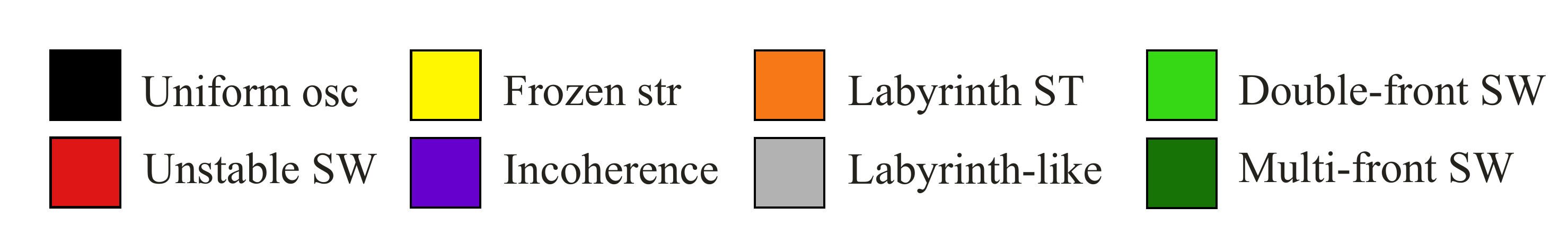}
\caption{(Color online) Regime maps of the network Eq.~\eqref{eq:grid} under external periodic extreme pulse force ((a) and (b)) and corresponding normalizing spectral density within the same ranges in $\omega$ ((c) and (d)). The black color is a regime of spatially uniform oscillations; the red color corresponds to unstable spiral waves; the yellow color is  ''frozen'' structures; the violet color is an incoherence regime; the orange color is labyrinth-like structures of a spiral type, the gray color is labyrinth-like structures, the light-green color corresponds to double-front spiral waves, dark-green color represents multi-front spiral waves.
Parameters: $\sigma=0.145 $ and $P = 2$, $a=1,\,b=3,\,c=1,\,d=5,\,r=0.006,\,s=4,\,\chi=1.6 $, $N=50$.} 
\label{fig:regime_map}
\end{figure}
We now study the dynamics of the system \eqref{eq:grid} under the external periodic pulse force \eqref{eq:Gauss}. We vary values of the external force parameters $A$ and $\omega$ within the following parameter ranges: $A \in [2.5,10]$ and $\omega \in[1.4,11]$.	
We are focusing on the impact of a strong external force, where the pulse's amplitude $A$ exceeds the amplitude of oscillations in the autonomous network. We select a specific range of pulse frequencies that align with the first four peaks in the spectrum of oscillations of the autonomous system (when $A$) (see Fig.~\ref{fig:initial_regime}(c)).  

In this work, we pay main attention to extreme amplitudes under external influence and when it exceeds the amplitude of oscillations in the autonomous system \eqref{eq:grid}. We have calculated the system dynamics for the chosen ranges of $A$ and $\omega$ and have found 8 new dynamical regimes which are associated with the external force. The maps of these regimes within the parameters space of ($A,\,\omega$) are depicted in Fig.~\ref{fig:regime_map}(a) and (b).

The localization of regimes in the ($A,\,\omega$) parameter plane significantly depends on an interrelation of the pulse frequency with the spectrum of the autonomous network \eqref{eq:grid} ($A=0$). We plot this power spectral density in Fig.~\ref{fig:regime_map}(c) and (d) in the same scale as the regime maps for clarity. 

When the amplitude $A$ is low, and value $\omega$ is not close to the spectral maxima, the initial spiral wave chimera loses its stability, and new spiral waves and spiral wave chimeras occur, but they are also unstable. The spatiotemporal patterns constantly change over time. In the regime map in Fig.~\ref{fig:regime_map}, this region corresponds to the red color. It is the most extensive region. It is observed for all the values of amplitudes and frequencies when $\omega > 10$, $\omega \in [5.6, 9.5]$ and for the wide region when $A$ is low. 

If the pulse frequency $\omega$ is close to the frequency of the main harmonic (see Fig.~\ref{fig:initial_regime}(c)), then the external pulse begins to synchronize oscillations in the system. We observe a regime of uniform oscillations or a regime with a smooth spatial profile: the instantaneous phases of adjacent oscillators are very similar, but outlying nodes can have to distinguish instantaneous phases. This regime corresponds to the black color in Fig.~\ref{fig:regime_map}. This region is extended with an increase in the amplitude $A$. Note that the region of this regime is observed not only close to the frequency of the main spectral peak but also for another value of $A$ and $\omega$. However, the region width does not already depend on the amplitude $A$.

When the external pulse frequency is close to the second or fourth spectral peaks (see  Fig.~\ref{fig:initial_regime}(c)), we observe a regime where the spatial structure is frozen (frozen structures regime). In this regime, several clusters with similar instantaneous phases of oscillations in nodes can be seen. The spatial localization of clusters does not change in time, and any spatial changes are absent. The frequency of oscillations for all the nodes is the same. This regime is highlighted by yellow color in Fig.~\ref{fig:regime_map}. This region is extended with an increase in the amplitude $A$. At the same time, the regime is also observed for other values of the frequencies of $\omega$, for example, for $\omega \approx 1.85$ and $\omega \approx 3.4$. 

In run-out mode, when the frequency $\omega$ of an external force lies far from the spectral maxima and for sufficiently large amplitude $A$, the external influence leads to the regime of incoherence when the instantaneous phases are different, even for neighbors. This regime is denoted by a violet color in the regime map.

The next interesting regimes can be considered as sub-types of the previous regime; they also represent frozen spatial structures but have significantly more complex shapes of a spatial profile. They look like labyrinth-like structures with alternating thin strips, which can have a spiral-like shape (orange color region in the regime map) or cross vertical and horizontal strips (a gray color region in the regime map). Interestingly, these regimes are observed to be generally close to the boundaries between the two regimes. 

The last regimes induced by the Gaussian pulses are spiral wave regimes. Their features significantly change from the ones in the autonomous case. The main distinction is that they do not determine the temporal dynamics of the lattice \eqref{eq:grid}, and the wave always has several close wavefronts. We highlight them with light-green (double-front wave) and dark-green ( multi-front wave) colors. They are also usually observed in a thin region between two dynamical regimes. The largest region is observed in the left boundary of the frozen structure region for $\omega\in [4.4, 4.6]$.
\textcolor{black}{Deducing stability of spatiotemporal patterns in the network of oscillators is a challenging problem due to their very high dimensions. It is a promising research problem to consider the stability of various spatiotemporal structures in complex networks. A good starting point would be the works below, which considered stability analysis of spiral waves in various reaction-diffusion systems. In \cite{Amdjadi2010}, authors have proposed a new numerical method for investigating the stability of spiral waves in reaction-diffusion systems. The stability of spiral waves has been explored in the locally coupled Kuramoto model under the effect of noise \cite{yu2023spatial}. Using linear and first-order approximation, they concluded that the spiral wave pattern is stable within a certain range of noise. However, the Hindmarsh-Rose model is significantly more complex, and rigorously determining stability is a promising future research direction.}

We below describe in detail the features of the main dynamical regimes denoted in the regime map in Fig.~\ref{fig:regime_map}.

\subsection{Dynamical regimes}
\begin{figure}[!h]
\centering
\hspace{-3mm}\parbox[c]{.35\linewidth}{ 
  \includegraphics[width=\linewidth]{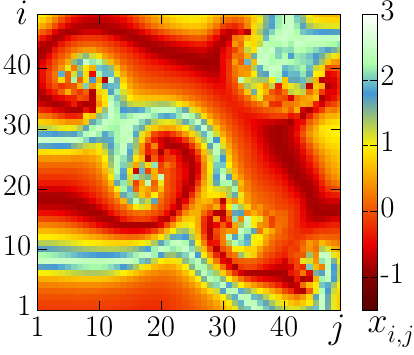}
    \vspace{-7.5mm} \center (a), $t=1$
     \includegraphics[width=\linewidth]{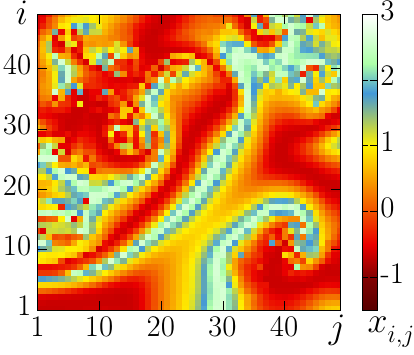}
    \vspace{-7.5mm} \center (c), $t=700$
}
\hspace{1mm}\parbox[c]{.35\linewidth}{
\includegraphics[width=\linewidth]{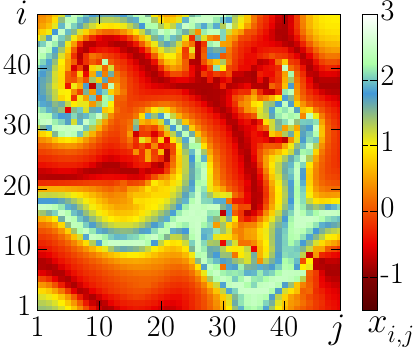}
\vspace{-7.5mm}\center (b), $t=200$
   \includegraphics[width=\linewidth]{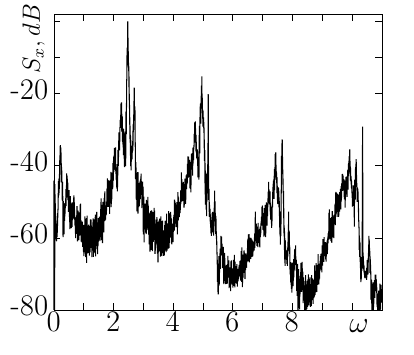}
    \vspace{-7.5mm} \center (d)
}

\caption{(Color online) Regime of unstable spiral waves (red color region in the regime map in fig.~\ref{fig:regime_map}) in the network Eq.~\eqref{eq:grid} under external force \eqref{eq:Gauss} with $A=3$ and $\omega=4.4$. (a)-(c) are snapshots of the system state at $t=1,\, t=200,\, t=700$ t.u., respectively, (d) shows normalized spectral density for $i = 31,\, j=25$. Parameters: $a=1,\,b=3,\,c=1,\,d=5,\,r=0.006,\,s=4,\,\chi=1.6 $, $N=50$.} 
\label{fig:unstable}
\end{figure}
\subsubsection{Unstable spiral waves}
We now study the regime of unstable spiral waves (red region in the regime map in Fig.~\ref{fig:regime_map}). Fig.~\ref{fig:unstable}(a)-(c) demonstrates three snapshots of the system state for the time moments $t = 1$, $t = 200$, and $t = 700$ t.u., respectively. The initial regime has already disappeared, and several new spiral wave chimeras are in the lattice \eqref{eq:grid}. However, localizations of the wave centers are shifted in times, as shown in Fig.~\ref{fig:unstable}(b). During this time, these waves are destroyed, and new spiral wave chimeras occur (Fig.~\ref{fig:unstable}(c)). They are also unstable. Thus,  the network has constant spatiotemporal variability when one wave emerges and destroys another, and so on. The spectrum is shown in Fig.~\ref{fig:unstable}(d) and is characterized by the same spectral peaks as in the initial spiral wave chimera (Fig.~\ref{fig:initial_regime}(c)) but becomes as noisy. Moreover, a peak at a very low frequency appears. It is absent in the spectrum of autonomous systems.

\subsubsection{Uniform oscillations}
\begin{figure}[!t]
\centering
\hspace{-3mm}\parbox[c]{.35\linewidth}{ 
  \includegraphics[width=\linewidth]{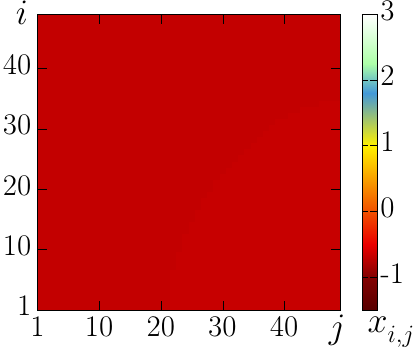}
    \vspace{-7.5mm} \center (a), $t=5$
     \includegraphics[width=\linewidth]{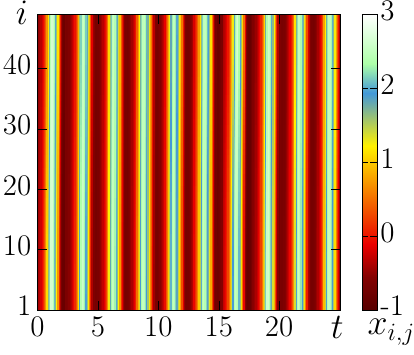}
    \vspace{-7.5mm} \center (c), $j=27$
}
\hspace{1mm}\parbox[c]{.35\linewidth}{
\includegraphics[width=\linewidth]{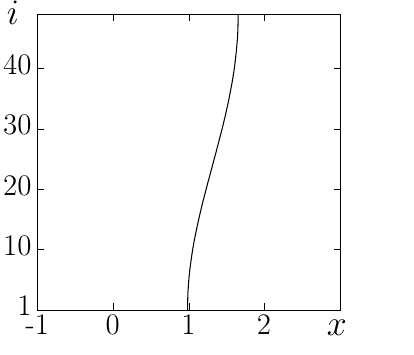}
\vspace{-7.5mm}\center (b), $j=27,\,t=500$
   \includegraphics[width=\linewidth]{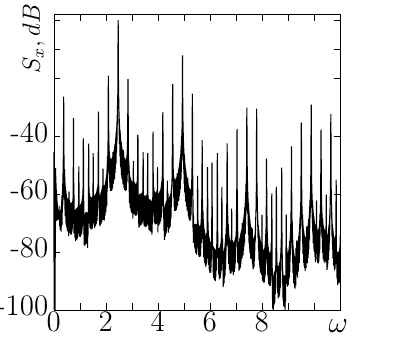}
    \vspace{-7.5mm} \center (d)
}

\caption{(Color online) Regime of spatially uniform oscillations (black color region in the regime map in fig.~\ref{fig:regime_map}) in the network Eq.~\eqref{eq:grid} under external force \eqref{eq:Gauss} with $A=3.6$ and $\omega=5.32$. (a) is a snapshot of the system state at $t=5$ t.u., (b) is an instantaneous spatial profile for cross-section $j=27$ at $t=500$, (c) is a space-time plot for cross-section $j=27$ (d) is normalized spectral density for $i = 25,\, j=25$. Parameters: $a=1,\,b=3,\,c=1,\,d=5,\,r=0.006,\,s=4,\,\chi=1.6 $, $N=50$.} 
\label{fig:uniform}
\end{figure}

In the regime map illustrated in Fig.~\ref{fig:regime_map}, the regime of uniform oscillations is one of the largest regions (black color). The initial regime completely disappears. The external pulse synchronises oscillations in the whole system, as is seen in Fig.~\ref{fig:uniform}(a). Every node in the system exhibits oscillations that are vastly similar in instantaneous phases, albeit not identical. This phenomenon is captured in the instantaneous spatial profile of a particular cross-section presented in Fig.~\ref{fig:uniform}(b), where values of $x_{i,j}$ variables are slightly different. At the same time, this difference changes in time from zero to  $\approx 0.8$. The spatiotemporal behavior for a chosen cross-section is depicted in a space-time diagram in Fig.~\ref{fig:uniform}(c). The regime is stable in time and is observed during the whole time of observation without any changes. The power spectral density is shown in Fig.~\ref{fig:uniform}(d). Interestingly, despite the quite close amplitudes and frequencies in this and previous regimes, their spectral distributions are significantly different. A lot of new harmonics appear in this case. \textcolor{black}{Such synchronized oscillations could potentially be utilized in a study of neural networks \cite{Tonnelier1999}, cardiac rhythms \cite{Goldbeter2022}, or other biological systems where coordinated activity is important}.

\begin{figure}[!t]
\centering
\hspace{-3mm}\parbox[c]{.35\linewidth}{ 
  \includegraphics[width=\linewidth]{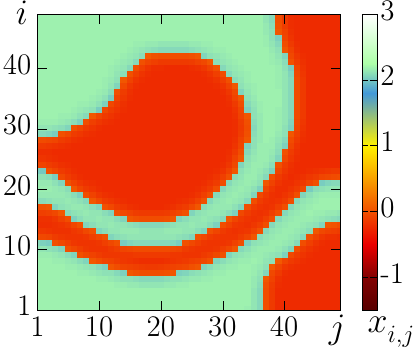}
    \vspace{-7.5mm} \center (a), $t=100$
     \includegraphics[width=\linewidth]{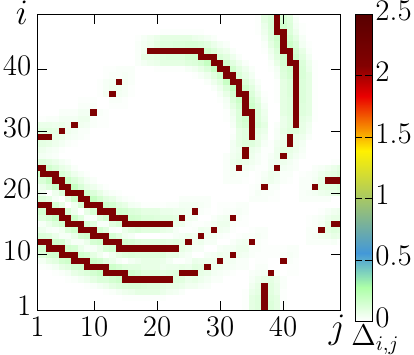}
    \vspace{-7.5mm} \center (c)
}
\hspace{1mm}\parbox[c]{.35\linewidth}{
\includegraphics[width=\linewidth]{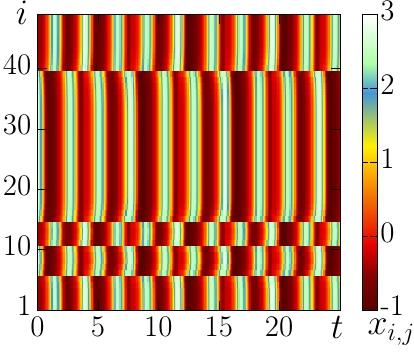}
\vspace{-7.5mm}\center (b), $j=16$
   \includegraphics[width=\linewidth]{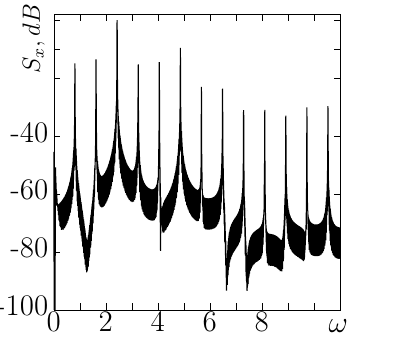}
    \vspace{-7.5mm} \center (d)
}

\caption{(Color online) Regime of frozen structures (yellow color region in the regime map in fig.~\ref{fig:regime_map}) in the network Eq.~\eqref{eq:grid} under external force \eqref{eq:Gauss} with $A=3.7$ and $\omega=1.62$. (a) is a snapshot of the system state at $t=100$ t.u., (b) is a space-time plot for cross-section $j=16$, (c) is the spatial distribution of RMSD, (d) shows normalized spectral density for $i = 31,\, j=25$. Parameters: $a=1,\,b=3,\,c=1,\,d=5,\,r=0.006,\,s=4,\,\chi=1.6 $, $N=50$.} 
\label{fig:frozen}
\end{figure}

\subsubsection{Frozen structures}
When the pulse frequency $\omega$ is close to the main spectral peak for the autonomous system \eqref{eq:grid} we observe a regime of frozen structures (highlighted in yellow in the regime map in Fig.\ref{fig:regime_map}. Apparently, it appears for the case when the system is most sensitive to external influence since the pulse frequency is close to the mean frequency. At the same time, the regime is also observed for other values of $\omega$, for example, when $\omega$ is close to the fourth spectral peak or even for values far from the spectral maxima. The regime is characterized by two-level clustering of the neurons. An example of this structure is depicted with a snapshot of the system state in 
\begin{figure}[!t]
\centering
\hspace{-3mm}\parbox[c]{.33\linewidth}{ 
  \includegraphics[width=\linewidth]{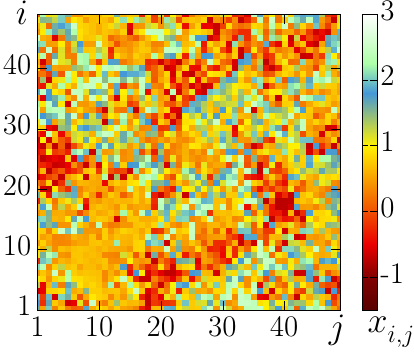}
    \vspace{-7.5mm} \center (a), $t=100$
}
\hspace{1mm}\parbox[c]{.33\linewidth}{
\includegraphics[width=\linewidth]{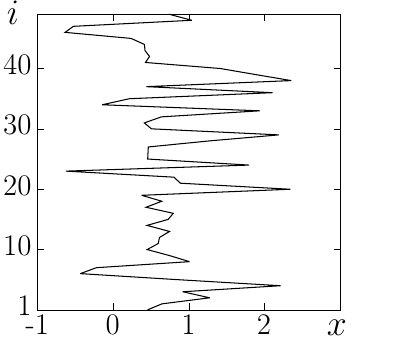}
\vspace{-7.5mm}\center (b), $j=16,\,t=100$
}
\hspace{-4mm}\parbox[c]{.33\linewidth}{
\includegraphics[width=\linewidth]{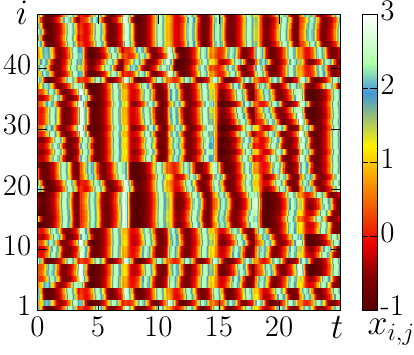}
\vspace{-7.5mm}\center (c), $j=16$
}
\caption{(Color online) Regime of incoherence (violet color region in the regime map in fig.~\ref{fig:regime_map}) in the network Eq.~\eqref{eq:grid} under external force \eqref{eq:Gauss} with $A=5.8$ and $\omega=1.74$. (a) is a snapshot of the system state, (b) is an instantaneous spatial profile for cross-section $j=24$ at $t=100$, (c) is a space-time plot for cross-section $j=24$. Parameters: $a=1,\,b=3,\,c=1,\,d=5,\,r=0.006,\,s=4,\,\chi=1.6 $, $N=50$.} 
\label{fig:incoherence}
\end{figure}
\begin{figure}[!t]
\centering
\hspace{-3mm}\parbox[c]{.33\linewidth}{
\center Snapshots
  \includegraphics[width=\linewidth]{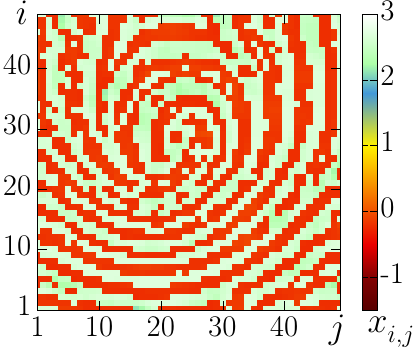}
    \vspace{-7.5mm} \center (a1), $A=3.8,\,\omega=0.85$
     \includegraphics[width=\linewidth]{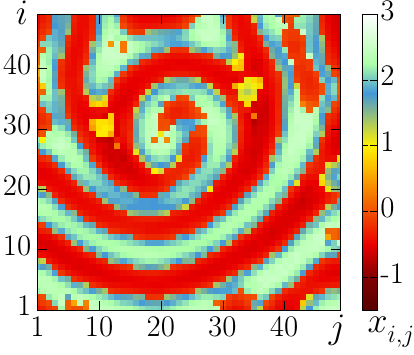}
    \vspace{-7.5mm} \center (a2), $A=3,\,\omega=2.53$
}
\hspace{1mm}\parbox[c]{.33\linewidth}{
\center Space-time plot
\includegraphics[width=\linewidth]{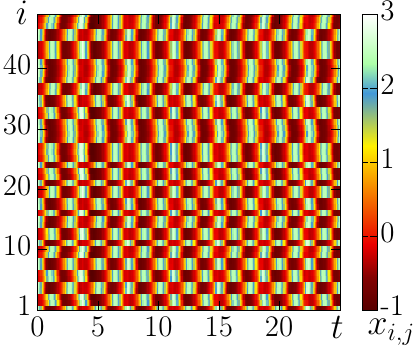}
\vspace{-7.5mm}\center (b1), $j=27$
   \includegraphics[width=\linewidth]{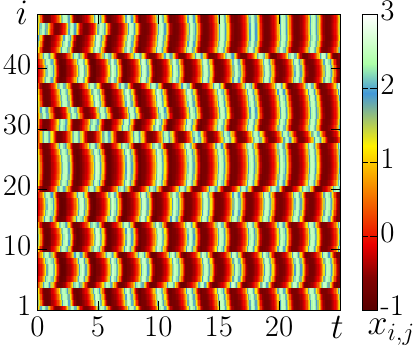}
    \vspace{-7.5mm} \center (b2), $j=29$
}
\hspace{1mm}\parbox[c]{.33\linewidth}{
\center Time series
\includegraphics[width=\linewidth]{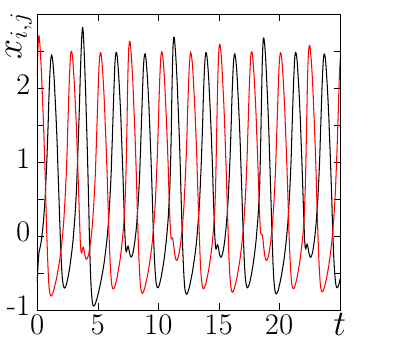}
\vspace{-7.5mm}\center (c1)
   \includegraphics[width=\linewidth]{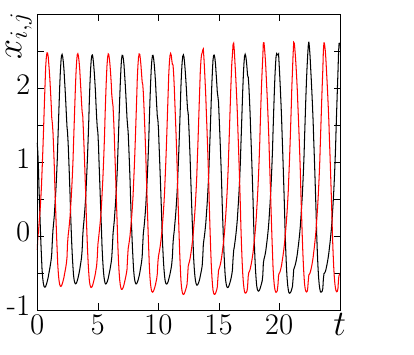}
    \vspace{-7.5mm} \center (c2)
}
\caption{(Color online) Regime of spiral-like structures (orange color region in the regime map in fig.~\ref{fig:regime_map}) in the network Eq.~\eqref{eq:grid} under external force \eqref{eq:Gauss} with $A=3.8$ and $\omega=1.7$ (top row) and $A=3.$ and $\omega=5.06$ (bottom row). (a1) and (a2)  are snapshots of the system state, (b1) and (b2) are space-time plots for cross-section $j=27$ and $j=29$, respectively, (c1) and (c2) time series for two elements from adjacent ''strips'' (c1: $i = 33,\, j=26$, black line and $i = 31,\, j=26$, red line) and (c2: $i = 31,\, j=26$, black line and $i = 33,\, j=26$, red line). Parameters: $a=1,\,b=3,\,c=1,\,d=5,\,r=0.006,\,s=4,\,\chi=1.6 $, $N=50$.} 
\label{fig:spiral_structure}
\end{figure}
Fig.~\ref{fig:frozen}(a). It is seen that the structure shape is similar to the initial spiral wave. However, any motion here is absent. The boundaries of the clusters are stable in time and do not change. In most cases, the structure forms by pausing the wavefront at a specific moment. Then, the spatial structure is ''frozen''. Note that the term ``frozen'' does not mean an absence of oscillations. If we consider the spatiotemporal dynamics in a space-time plot in Fig.~\ref{fig:frozen}(b), then we can see that it is a standing wave. Each element oscillates in time with the same frequency. Oscillators inside one cluster have almost the same instantaneous phases, while oscillators from adjacent clusters have significantly different phases. It is clearly seen in the spatial distribution of the RMSD $\Delta_{i,j}$ in Fig.~\ref{fig:frozen}(c). Values of $\Delta_{i,j}$ are very low inside the clusters and large at the boundaries of clusters. The spectrum of oscillation shown in Fig.~\ref{fig:frozen}(d) is smoother than for the previous regimes. It always has two additional harmonics between the initial spectral maxima, which is associated with the external force. The regime of frozen oscillation is typical for biological systems. \textcolor{black}{For example, frozen structures in cardiac rhythms can manifest as sustained patterns of heart muscle contractions that persist unchanged over a specific time interval \cite{Esperer2008}}.

\subsubsection{Incoherence regime}
Another destructive behavior of pulses, except for the regime of the unstable spiral wave, is observed for the high amplitude of pulses $A$ with a frequency far from the spectral maxima (violet regions in the regime map in Fig.~\ref{fig:regime_map}). In this case, the neurons begin to oscillate incoherently. Note that a similar regime is possible only for completely almost uncoupled autonomous networks. This regime is represented in Fig.~\ref{fig:incoherence}(a). Instantaneous states for the neurons are not significantly different, as shown in an instantaneous spatial cross-section in Fig.~\ref{fig:incoherence}(b). The spatiotemporal dynamics are sufficiently complex. It is demonstrated by a space-time plot of the cross-section in Fig.~\ref{fig:incoherence}(b). One can see that a large part of neurons oscillate similarly but with a certain inhomogeneous delay. This effect looks like the strips at the same time (red or green). Moreover, unstable short-time clustering takes place in different parts of the network.

\subsubsection{Labyrinth-like structures}
Interesting behavior is observed close to boundaries (see in Fig.~\ref{fig:regime_map})  of two regimes described above, for example, between the frozen structure regime and unstable spiral waves. Here, we discover a new regime, which we call a labyrinth-like structure of spiral type. The regime has similar temporal dynamics as the frozen structures, but the spatial features are different. The spatial shape of a standing wave becomes more complicated, and it represents two thin spiralled strips. \textcolor{black}{The similar structures have been discovered in another system of coupled oscillators, for example, in \cite{Luo2009, Dai2021}}. We find two types of these structures with strip widths equal to 2-4 nodes and wider strips with a width of 3-8 nodes. These are captured in the snapshots showcasing the system's states seen in Fig.~\ref{fig:spiral_structure}(a1) and (a2). Note that the strip width is not constant and can change along the spiral. Small clusters with another instantaneous phase can be located inside the strips. Despite the apparent similarity with a spiral wave, we do not observe any rotation of the fronts. The spatiotemporal dynamics are the same for both types of regimes. Oscillators inside one strip have similar instantaneous phases along the whole strip, which slightly change near the boundaries between two strips. It is depicted in space-time plots of a spatial cross-section through the spiral center in Fig.~\ref{fig:spiral_structure}(b1) and (b2). Oscillations in the adjacent strips have a phase difference close to $\pi/2$ as shown in the time series for selected elements in Fig.~\ref{fig:spiral_structure}(c1) and (c2).

\begin{figure}[!h]
\centering
\hspace{-3mm}\parbox[c]{.33\linewidth}{
\center Snapshots
  \includegraphics[width=\linewidth]{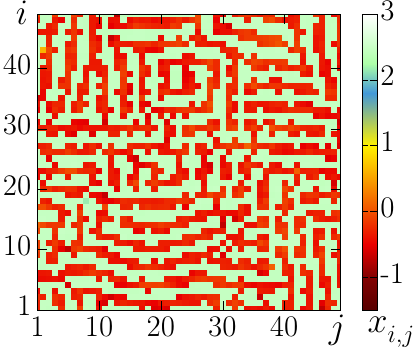}
    \vspace{-7.5mm} \center (a1), $A=7,\,\omega=0.85$
     \includegraphics[width=\linewidth]{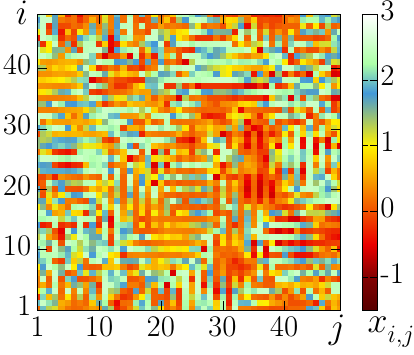}
    \vspace{-7.5mm} \center (a2), $A=3.9,\,\omega=1.04$
}
\hspace{1mm}\parbox[c]{.33\linewidth}{
\center Space-time plot
\includegraphics[width=\linewidth]{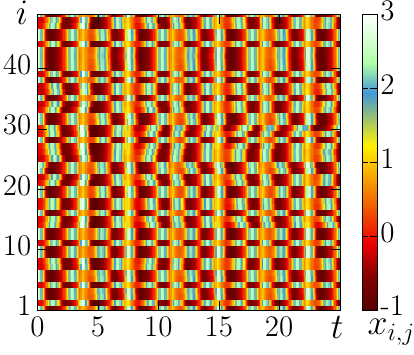}
\vspace{-7.5mm}\center (b1), $j=16$ 
   \includegraphics[width=\linewidth]{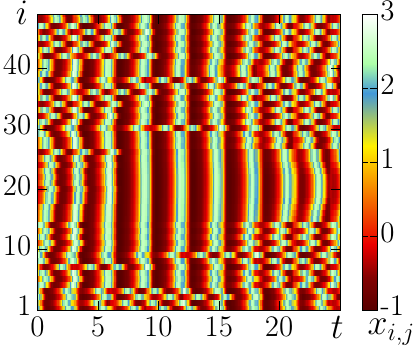}
    \vspace{-7.5mm} \center (b2),  $j=15$
}
\hspace{1mm}\parbox[c]{.33\linewidth}{
\center Time series
\includegraphics[width=\linewidth]{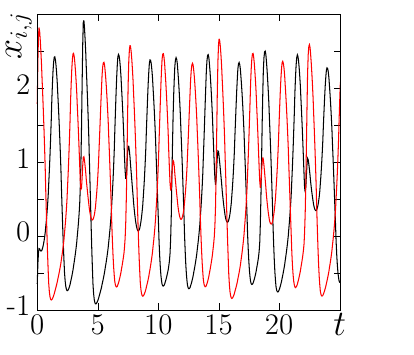}
\vspace{-7.5mm}\center (c1)
   \includegraphics[width=\linewidth]{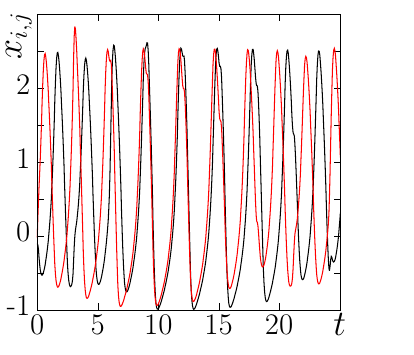}
    \vspace{-7.5mm} \center (c2)
}
\caption{(Color online) Regime of labyrinth-like structures (gray color region in the regime map in fig.~\ref{fig:regime_map}) in the network Eq.~\eqref{eq:grid} under external force \eqref{eq:Gauss} with $A=7$ and $\omega=1.7$ (top row) and $A=3.9$ and $\omega=2.08$ (bottom row). (a1) and (a2)  are snapshots of the system state, (b1) and (b2) are space-time plots for cross-section $j=16$ and $j=15$, respectively, (c1) and (c2) time series for two elements from adjacent ''strips'' (c1: $i = 33,\, j=25$, black line and $i = 35,\, j=25$, red line) and (c2: $i = 32,\, j=24$, black line and $i = 33,\, j=24$, red line). Parameters: $a=1,\,b=3,\,c=1,\,d=5,\,r=0.006,\,s=4,\,\chi=1.6 $, $N=50$.} 
\label{fig:labyrinth-like}
\end{figure}

There is a very similar regime to the previous one, which is observed within the similar values of $A$ and $\omega$ inside the gray region of the regime map in Fig.~\ref{fig:regime_map}. The spatial profile also represents an alternation of the thin strips, but they are located in general along vertical or horizontal directions, intersecting with each other. We call this regime a labyrinth-like structure. \textcolor{black}{Similar patterns have been observed in another network of coupled neurons in \cite{Feng2021}}. Two typical cases of them are depicted in Fig.~\ref{fig:labyrinth-like}(a1) and (a2). The strip width can be 2-3 nodes (a1) or only 1 node (a2). We have discovered similar structures in the lattice of van der Pol oscillators with either strong coupling or repulsive coupling \cite{shepelev2021bistable, shepelev2021repulsive}. However, thin strips have been observed only for local coupling, while wide ones have been for nonlocal interaction. The spatiotemporal dynamics for both types are shown by space-time plots of cross-sections in Fig.~\ref{fig:labyrinth-like}(b1) and (b2). For the first case, the spatial structure is stable in time; all oscillators of one strip have similar instantaneous phases, while elements of adjacent strip oscillate almost in anti-phase. This is shown in the time series for the two chosen nodes from adjacent strips in Fig.~\ref{fig:labyrinth-like}(c1). The spatial profile of the second type of labyrinth-like structure is unstable. The strips can be realigned when new intersections of horizontal and vertical strips occur and disappear. This process is illustrated in the space-time plot in Fig.~\ref{fig:labyrinth-like}(b2) as well as in the time series in Fig.~\ref{fig:labyrinth-like}(c2). At the beginning, the two nodes oscillate almost in anti-phase. However, their strips lose stability, and after this, both nodes are in the same strip. Their instantaneous phases are very similar. But they begin to oscillate in anti-phase again after subsequent restructuring.

\begin{figure}[!t]
\centering
\hspace{-3mm}\parbox[c]{.33\linewidth}{ 
  \includegraphics[width=\linewidth]{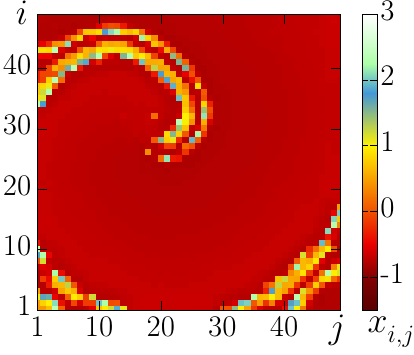}
    \vspace{-7.5mm} \center (a1), $t=20$
     \includegraphics[width=\linewidth]{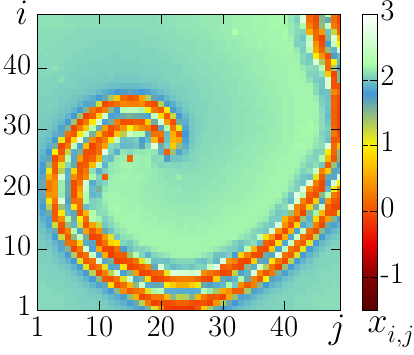}
    \vspace{-7.5mm} \center (a2), $t=360$
}
\hspace{1mm}\parbox[c]{.33\linewidth}{
\includegraphics[width=\linewidth]{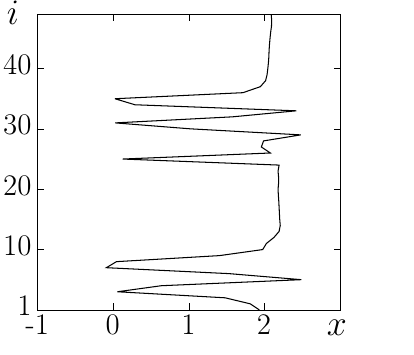}
\vspace{-7.5mm}\center (b), $t=360,\, j=16$
   \includegraphics[width=\linewidth]{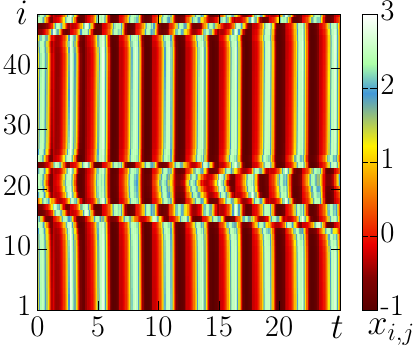}
    \vspace{-7.5mm} \center (d) $j=15$
}
\hspace{1mm}\parbox[c]{.33\linewidth}{
\includegraphics[width=\linewidth]{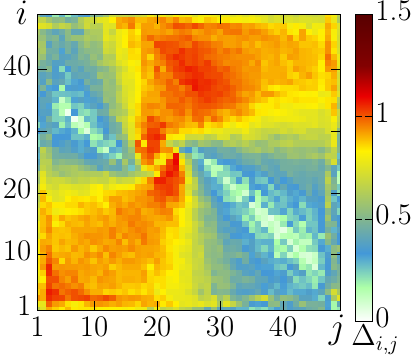}
\vspace{-7.5mm}\center (c)
   \includegraphics[width=\linewidth]{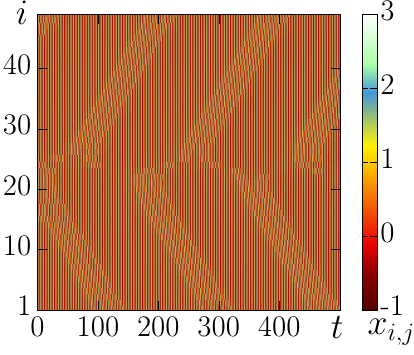}
    \vspace{-7.5mm} \center (e) $j=15$
}
\caption{(Color online) Regime of double-front spiral wave (light-green color region on the regime map in fig.~\ref{fig:regime_map}) in the network Eq.~\eqref{eq:grid} under external force \eqref{eq:Gauss} with $A=9.9$ and $\omega=5.04$. (a1) and (a2)  are snapshots of the system state at $t=20$ and $t=360$, respectively, (b) is an instantaneous spatial profile for the cross-section $j=15$, (c) is a spatial distribution of RMSD, (d) and (e) show space-time plots for cross-section $j=15$ for short and long time periods, respectively, Parameters: $a=1,\,b=3,\,c=1,\,d=5,\,r=0.006,\,s=4,\,\chi=1.6 $, $N=50$.} 
\label{fig:spiral_wave1}
\end{figure}
\begin{figure}[!t]
\centering
\hspace{-3mm}\parbox[c]{.33\linewidth}{ 
  \includegraphics[width=\linewidth]{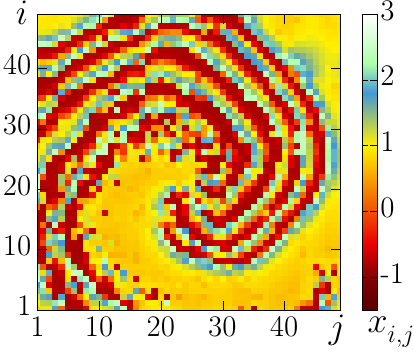}
    \vspace{-7.5mm} \center (a1), $t=1$
     \includegraphics[width=\linewidth]{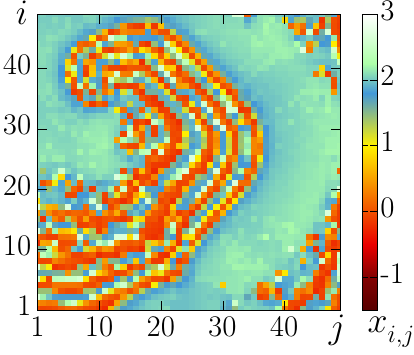}
    \vspace{-7.5mm} \center (a2), $t=440$
}
\hspace{1mm}\parbox[c]{.33\linewidth}{
\includegraphics[width=\linewidth]{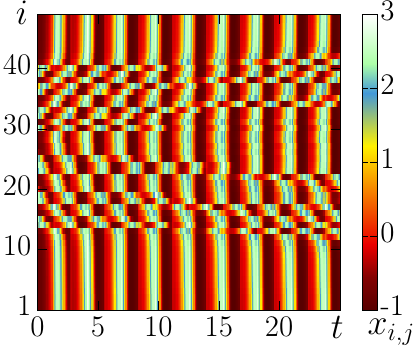}
\vspace{-7.5mm}\center (b), $j=12$
   \includegraphics[width=\linewidth]{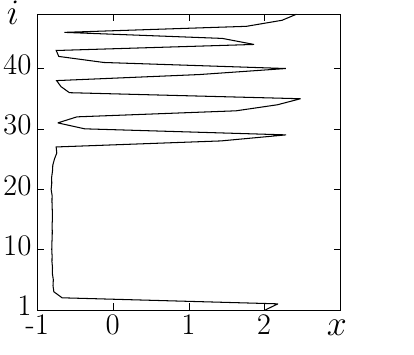}
    \vspace{-7.5mm} \center (d), $t=1,\,j=12$
}
\hspace{1mm}\parbox[c]{.33\linewidth}{
\includegraphics[width=\linewidth]{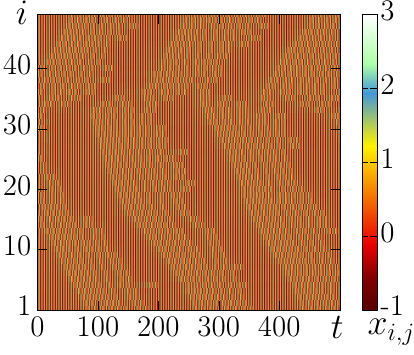}
\vspace{-7.5mm}\center (c), $j=12$
   \includegraphics[width=\linewidth]{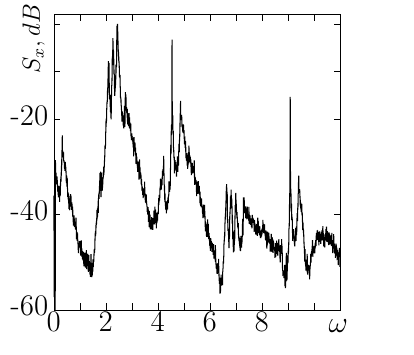}
    \vspace{-7.5mm} \center (e)
}
\caption{(Color online) Regime of multi-front spiral wave (dark-green color region in the regime map in fig.~\ref{fig:regime_map}) in the network Eq.~\eqref{eq:grid} under external force \eqref{eq:Gauss} with $A=9.9$ and $\omega=5.04$. (a1) and (a2)  are snapshots of the system state at $t=1$ and $t=440$, respectively; (b) and (c) show space-time plots for cross-section $j=12$ for short and long time periods, respectively; (d) is an instantaneous spatial profile for the cross-section $j=15$ at $t=1$, (e) is a normalized spectral density for $i = 33,\, j=24$. Parameters: $a=1,\,b=3,\,c=1,\,d=5,\,r=0.006,\,s=4,\,\chi=1.6 $, $N=50$.} 
\label{fig:spiral_wave2}
\end{figure}

\subsubsection{Multi-front spiral waves}

We now consider the two wave regimes, which are sufficiently rare but are the most interesting. They are observed in light-green and dark-green regions of the regime map shown in Fig.~\ref{fig:regime_map}. They are a new type of spiral wave, but their features are significantly different from common spiral waves in the autonomous system. Their occurrence is due to external pulse influences. For the first kind of these spiral waves, an illustration is provided through two system snapshots captured at moments $t=20$ and $t=360$, depicted in Fig.~\ref{fig:spiral_wave1}(a1) and (a2) respectively. A noteworthy observation is the behavior of the spiral wave's rotation; unlike the conventional single wavefront progression in typical spiral waves, here, a double wavefront is seen to propagate. This dual front's contour is visualized in an immediate spatial cross-section across the wave's center shown in Fig.~\ref{fig:spiral_wave1}(b) at $j=15$, revealing that the wavefront is composed of two ''folds''. The regime's RMSD spatial distribution is displayed in Fig.~\ref{fig:spiral_wave1}(c), where it's common to find the highest $\Delta_{i,j}$ values around the wave center. However, if we consider the wave dynamics in more detail, we may see that the doubled wavefront is not the only difference. The spatiotemporal dynamics are represented in a space-time diagram for the same cross-section, as in Fig.~\ref{fig:spiral_wave1}(d). Interestingly, the network dynamics are not determined by the wave. All the network nodes oscillate similarly to the previous regimes independently of the wave rotation. Only a sharp jump of instantaneous phases takes place for the wavefront oscillators. Even the wavefront elements oscillate in time. Moreover, noting wave motion is not observed in Fig.~\ref{fig:spiral_wave1}(d). Yet, when examining the dynamics over an extended period, as shown in a space-time plot in Fig.~\ref{fig:spiral_wave1}(e), a very slow movement of the spiral wave can be observed, manifested as moving the phase shifts. Indeed, the rotation velocity is many times slower than for the initial spiral wave in the autonomous network \eqref{eq:grid}.

In the regime map shown in Fig.~\ref{fig:regime_map}, the second type of the spiral wave (dark-green region) is illustrated by two snapshots of the system state at $t=1$ and $t=440$ t.u. in Fig.~\ref{fig:spiral_wave2}(a1) and (a2). Unlike the first type of spiral wave, this wavefront is not doubled but consists of several fronts (3-5 nodes) and slowly rotates around the wave center. However, this spatiotemporal structure is unstable in time. When observing the cross-section at the wave's core (Fig.~\ref{fig:spiral_wave2}(b)), the boundary of the front is unstable, with the edge fronts, especially from the inner side of the wave, being destroyed and occurring again during the rotation. Despite these differences, the features are very similar to the first type of spiral wave, as all the elements oscillate independently, and the wave velocity is also very similar. The power spectrum density in Fig.~\ref{fig:spiral_wave2}(e) shows that the spectrum does not explicitly include the harmonics of the pulse signal but consists of four noisy spectral peaks of the initial regime and a new low-frequency peak. In general, the spectral density is similar to the regime of unstable spiral waves (Fig.~\ref{fig:unstable}(d)).

Consequently, the significant external influence by Gaussian pulses \eqref{fig:GaussianForce} results in the emergence of various new dynamic regimes which are never observed in the autonomous network described by \eqref{eq:grid}. This influence can be both destructive and constructive. The destructive role is associated with losing the stability of initial spiral waves and the regime of incoherent oscillations. On the other hand, the external force brings about the appearance of consistent oscillations, stationary patterns, two different types of labyrinth-like structures, and two variations of spiral waves displaying new distinctive characteristics.

\section*{Conclusion}
We have studied numerically the effects of the external influence by the periodic high-amplitude Gaussian pulses on the network of nonlocally coupled Hindmarsh-Rose neurons with 2D geometry of the coupling topology. The boundary conditions correspond to the zero flux. The strong amplitude of pulses means that it is larger than the amplitude of oscillations in the autonomous network; namely, we consider the influence with amplitude $A\in [2.5,\,10]$. The initial regime in the network is a regime of the spiral wave chimera. We have calculated and plotted a regime map on the plane defined by the external force parameters $(A,\,\omega)$, where each regime is represented by a distinct color. We have discovered that the external pulse force can play both destructive and constructive roles, namely, destroy the initial regime or lead to the emergence of new types of regimes, which are never observed in the autonomous system.

Our research has demonstrated that a weak external influence can cause the initial spiral wave chimera to lose stability. After a certain time, the initial spiral wave chimera loses its stability and is replaced by a new unstable spiral wave or multiple unstable waves in the network. These new waves exist for a short period before being replaced by another set. This behavior is observed for any pulse frequencies larger than $\omega > 10$ for any amplitude, as well as for lower frequencies at low amplitude. It is the most expected effect of the pulse influence.

Another type of destructive impact of pulses is an occurrence of completely incoherent behaviour. It is observed when the amplitude of an external force is large, and the frequency is far from the frequencies of spectral peaks. In this case, the oscillations is similar to ones in the uncoupled network.

When we impact by pulses with a frequency ($\omega$) close to the main spectral peak, we observe the first constructive effect of the external force. The pulses synchronize oscillations of all the network nodes, and the system displays uniform oscillations, when all the oscillators behave in the same way. However, when the influence frequency $\omega$ is close to the other spectral peak frequencies, the behavior is different from the previous case. A new spatiotemporal structure emerges in the network. It represents a cluster structure with two distinct levels, inside which the instantaneous phases of oscillations are similar, while neurons of different levels oscillate to anti-phase. A spatial shape has certain features of the initial regime. This regime is typical in different oscillatory networks.

A deviation of the pulse frequency from the spectral peaks leads to the restructuring of the cluster spatiotemporal structure. The spatial structure becomes spiral-shaped and irregular in space; however, it saves the cluster character with two levels. The spatial shape represents irregular alternations of thin and sufficiently long regions (strips). For this reason, we have called this regime as the labyrinth-like structure of a spiral type. A following deviation of the pulse frequency leads to a change in the type of the cluster structure. The ''strips'' become not spiral but vertical or horizontal with complex shapes. The spatial structure represents the alternation or intersections of these ''strips''. These types of spatiotemporal structures have also been discovered in other oscillatory networks, including neural ones.

The most interesting response to the external pulses is observed on the boundary of the parameter region between the regimes described above. It is a regime of \textit{new} type of spiral waves. For this regime, oscillations of all the neurons are independent of the wave dynamics, while only the phase shift is observed along the wavefront. Moreover, the wave rotation is accompanied by a motion of the several (from 2 to 5) wavefronts located close to each other. The wave velocity is very slow; it is many times lower than for the spiral wave in the autonomous system. \textcolor{black}{At the same time, such waves are not completely stable. The width of their wavefronts can change in time within a certain range, and some fronts can also change in time, i.e., some fronts appear and disappear during the observation}. This type of spiral wave has not been found in other oscillatory networks and, apparently, becomes possible due to regular pulse impact. A question arises, does this type of wave occur only for this type of influence or for other types, too?

Thus, the external force of periodic Gaussian pulse series with strong amplitudes can lead not only to destructive behavior but also to several new regimes that remain stable during the whole observation time and are never observed in the autonomous system. Our calculations show a crucial role in the frequency of pulses in conjunction with their amplitude. Note that most of the regimes induced by the external force are impossible in the autonomous network, but they are observed in other oscillatory and neural networks and can play an important role in their dynamics. Hence, the influence of high-amplitude periodic pulse is an effective approach to control the network dynamics, which enables us to induce the new dynamic regimes that we need. The following object of the study is discovering methods for more precise control of parameters, the contribution of the network topology and the effect of irregular pulse series with both normal and extreme amplitudes.

\section*{Acknowledgments}
The study was supported by the Russian Science Foundation (project No. 23-72-10040, \url{https://rscf.ru/en/project/23-72-10040/}). The authors express gratitude for the NVIDIA DGX cluster provided by Digital University Kerala, as well as for the additional computing resources made available by the University. The authors would like to thank Prof. Tatyana E. Vadivasova for her help in understanding the results.

\section*{Conflict of interest}
 The authors declare that they have no conflict of interest.

\section*{Data Availability Statement}
The data that support the findings of this study are available within the article.

\section*{Credit author statement}
\textbf{J.S. Ram}: Writing: writing: review $\&$ editing, investigation, analysis
\newline
\textbf{S.S. Muni}: Supervision, writing: review $\&$ editing, investigation, formal analysis
\newline
\textbf{I.A. Shepelev}: Supervision, Writing: original draft, software, investigation, analysis, visualization

\bibliographystyle{unsrt} 
\bibliography{Arxiv}
\end{document}